 \documentstyle[12pt,aps,prb]{revtex}

\begin{document}
{\pagestyle{empty}
\rightline{MTU-PHY-HA98/5}
\rightline{July 1998}
\vskip 2.5cm
{\renewcommand{\thefootnote}{\fnsymbol{footnote}}
\centerline{\large \bf Finite-size scaling } 
 \vskip 0.25cm
 \centerline{\large \bf of  helix-coil transitions in poly-alanine}
\vskip 0.25cm
\centerline{\large \bf studied by multicanonical simulations}
}
\vskip 2.5cm
 
\centerline{Ulrich H.E.~Hansmann$^{\#,}$\footnote{hansmann@mtu.edu; 
  to whom correspondence should be addressed.}
 and Yuko Okamoto$^{\ddagger,}$\footnote{okamotoy@ims.ac.jp}} 
\vskip 1.0cm

\centerline{$^\#${\it Dept.~of Physics, Michigan Technological University,
 Houghton, MI 49931-1291, USA}} 
\vskip 0.25cm
\centerline {$^\ddagger${\it Dept.~of  
Theoretical Studies, Institute for Molecular Science, Okazaki, Aichi 444-8585,
Japan}}
 
\medbreak
\vskip 3.0cm
 
\centerline{\bf ABSTRACT}
We report results from multicanonical simulations of poly-alanine.
Homopolymers of up to 30 amino acids were considered and various
thermodynamic quantities as a function of temperature calculated. 
We study the nature of the observed helix-coil transition and 
present estimates for critical exponents.
\vfill
\newpage}
 \baselineskip=0.8cm
\noindent
{\bf INTRODUCTION} \\
It is widely believed that the folding of a protein into its native
structure involves one or more transitions between distinct
phases. However,  these transitions are not fully
understood. Are they simply crossovers between conformers   or is 
it justified to refer to them 
(as commonly done) as phase transitions? Strictly speaking, the latter 
concept is only well defined for infinite  systems. It is not clear whether
proteins can be regarded as ``almost'' infinitely large systems (in the sense 
in which 
the $10^{23}$ atoms in a crystal form an ``infinite'' system). The 
properties of proteins 
depend strongly
on  the number and composition of amino acids and  therefore follow {\it not}
from the collective behavior of a great many similar unit
objects. For the same reason, it is also not possible to study the transition
by considering  extrapolation of a given protein to an infinitely 
long chain: 
The properties of the protein may change completely by adding
or substracting an amino acid. Hence, finite-size scaling analyses, 
a common tool to
study phase transitions, is not applicable. 
The exception are  homopolymers of amino acids.
Here, an extrapolation to infinitely long chains is possible and it 
should indeed be 
possible to describe  
for these molecules configurational changes 
within the framework of  phase transitions. 
Hence,  homopolymers of amino acids are suitable models for 
investigation of how the observed transitions are affected by
the finite size of the molecule. 

In an earlier work \cite{OH95a} we studied  for three
characteristic amino acids (alanine, valine, and glycine)
$\alpha$-helix formation in short peptide systems and
compared our results with that of recent experiments 
(which are reviewed in Ref.~2). The helix-coil transition
was also studied
theoretically by various other 
groups  \cite{OO,TRJ,TB,DKK,VWGWS,CR,YOP,ZKWH,Jeff,GK}. 
For poly-alanine we observed 
a sharp transition between  disordered coil conformers and an ordered
phase where the polymer is in a helical state. It is tempting to describe
the poly-alanine chain in the framework of  Zimm-Bragg-type
theories \cite{ZB} in which the homopolymers  are described by a
one-dimensional 
Ising model  with the residues  as ``spins'' taking values 
``helix'' or ``coil''. By construction, the length of helical segments cannot
grow indefinitely in such models, since long-range order is not possible
for a one-dimensional system with solely short-range interactions. Hence,
 phase transitions are not possible in Zimm-Bragg model. However, such models 
may be  too crude a description.
In the helix-coil transition the polymer chain changes from a disordered
three-dimensional coil to an ordered and only quasi one-dimensional helix. This
is certainly different from spin-flips in a one-dimensional chain of
Ising spins.  In addition, 
the amino-acid residues in a homopolymer
are also subject to long-range (electrostatic) 
interactions which  allow even for  the case of a
 one-dimensional spin chain  long-range order and phase transitions.
Hence, it is an open  question whether  the observed helix-coil transition 
in poly-alanine is a phase transition or whether 
it is rather  a crossover between the two states. In order to find an answer 
to this question we can study finite chains and extrapolate the results to the 
limit of an infinitely long polymer. The limitations of our previous results
did not allow such analyses.
Here, we report results from a simulation of poly-alanine with both increased
statistics and larger chains.

As in the previous article, the use of the multicanonical algorithm \cite{MU} 
was crucial. The various competing interactions within the polymer lead
to an energy landscape characterized by a multitude of local minima.
Hence, in the low-temperature region, canonical Monte Carlo or
molecular dynamics simulations will tend to get trapped in one of these
minima and the simulation will not thermalize within the available
CPU time. To overcome this problem we have proposed the application
of the multicanonical algorithm \cite{MU} and other {\it generalized-ensemble}
techniques \cite{Review} to the protein folding problem \cite{HO}.   We could
demonstrate that these techniques are indeed superior to standard
methods \cite{HO94c,HO96b} and  a useful tool for investigations of the
proteins and peptides, \cite{EH96d,HMO97b,HO97d} since they 
allow not only to find the ground-state conformer 
 but also to calculate thermodynamic
quantities at various temperatures from one simulation run. In the
latter point
our approach differs from global optimization methods which
were also used in previous works of protein folding. \cite{WC,FL}

Here we use the multicanonical algorithm to calculate various 
thermodynamic quantities  as a function of temperature
for poly-alanine of four different chain lengths. We concentrate on    
such quantities (like average number of helical residues or specific 
heat) where we expect to see the strongest signal for the helix-coil 
transition.  The finite-size scaling of these quantities is studied
and estimates for critical exponents are calculated. This allows us to
investigate the nature of the phase transition in poly-alanine. 
Finally, we update our results on the Zimm--Bragg parameters $s$ and
$\sigma$.
\\

\noindent
{\bf METHODS} \\
\noindent
{\bf Peptide Preparation and Potential Energy Function} \\
We considered amino-acid homo-oligomers of alanine which is known 
from experiments and our previous work to be 
a strong helix former at low temperatures. 
 The number of residues, $N$, was taken to be
10, 15, 20, and 30 in order to examine the $N$ dependence.
Since the charges at peptide termini are known to
reduce helix content,\cite{Ooi,SKY}
we removed them by taking a neutral NH$_2$-- group at the
N-terminus and a neutral --COOH group at the C-terminus. 

The potential energy function
$E_{tot}$ (in kcal/mol) that we used is given by the sum of
the electrostatic term $E_{C}$, 12-6 Lennard-Jones term $E_{LJ}$, and
hydrogen-bond term $E_{HB}$ for all pairs of atoms in the peptide together with
the torsion term $E_{tor}$ for all torsion angles:
\begin{eqnarray}
E_{tot} & = & E_{C} + E_{LJ} + E_{HB} + E_{tor},\\
E_{C}  & = & \sum_{(i,j)} \frac{332q_i q_j}{\epsilon r_{ij}},\\
E_{LJ} & = & \sum_{(i,j)} \left( \frac{A_{ij}}{r^{12}_{ij}}
                                - \frac{B_{ij}}{r^6_{ij}} \right),\\
E_{HB}  & = & \sum_{(i,j)} \left( \frac{C_{ij}}{r^{12}_{ij}}
                                - \frac{D_{ij}}{r^{10}_{ij}} \right),\\
E_{tor}& = & \sum_l U_l \left( 1 \pm \cos (n_l \chi_l ) \right).
\end{eqnarray}
Here, $r_{ij}$ (in \AA) is the distance between the atoms $i$ and $j$, and
$\chi_l$ is
the torsion angle for the chemical bond $l$.
The factor 332 in $E_C$ is to give the energy in units of
kcal/mol.
The parameters ($q_i,A_{ij},B_{ij},C_{ij},
D_{ij},U_l$, and $n_l$) for the energy function were adopted
from ECEPP/2.\cite{EC1,EC2,EC3} Note that in ECEPP/2 irrelevant
constant terms (that do not depend on conformations of the peptide)
are already subtracted from $E_{tot}$.  Since one can avoid the
complications of electrostatic and hydrogen-bond interactions of
side chains with the solvent for nonpolar amino acids, explicit
solvent molecules were neglected for simplicity 
and the dielectric constant $\epsilon$ was set equal to 2.
The computer code
KONF90 \cite{KONF} was used which differs slightly in 
conventions from the original version of ECEPP/2 (for example, 
$\phi_1$ of KONF90 is equal to
$\phi_1 - 180^{\circ}$ of ECEPP/2, and energies are also different by
small irrelevant constant terms). The peptide-bond
dihedral angles $\omega$ were fixed at the value 180$^\circ$
for simplicity,
which leaves $\phi_i,~\psi_i$, and $\chi_i$ ($i=1, \cdots, N$)
as independent degrees
of freedom.  Since alanine has only one $\chi$ angle in the side chain,
the numbers of independent degrees of freedom are $3 N$ where $N$ is
the number of residues.\\

\noindent
{\bf Simulation Techniques}

The Monte Carlo method that we used is the {\it multicanonical algorithm}
\cite{MU}, which is sometimes also
referred to  as {\it entropic sampling} \cite{ES} (in Ref.~31
it was shown that both algorithms are mathematically identical). Unlike 
in canonical simulations,
configurations with energy $E$ are here assigned a weight:
\begin{equation}
 w_{mu} (E)\propto \frac{1}{n(E)} = e^{-S(E)}~,
\label{eq2}
\end{equation}
where
\begin{equation}
S(E) = \log n(E)
\label{eq3}
\end{equation}
is the microcanonical entropy. A standard update scheme
such as  the Metropolis algorithm \cite{Metro} 
 will  realize a Markov chain in that ensemble and yield  a
 uniform distribution of energy:
\begin{equation}
 P_{mu}(E)
\propto n(E)~w_{mu}(E) = {\rm const}~.
\end{equation}
 All energies appear with  equal probability and 
a free random walk in the energy space is enforced. Hence, the
simulation can overcome any  energy barrier and will not get trapped in
one of the many local minima.
Since a large range of energies is sampled, the use of the
reweighting techniques
\cite{FS} allows us to calculate the expectation value of any physical
quantity ${\cal O}$ for a wide range of temperatures $T$ by
\begin{equation}
< {\cal O} >_T ~= \frac{\displaystyle{\int dE~ {\cal O} (E) 
                                      P_{mu} (E)~w^{-1}_{mu} (E)~
                                      e^{-{\beta} E}
                                     }}
                       {\displaystyle{\int dE~ P_{mu} (E)~w^{-1}_{mu} 
                                      (E)~e^{-{\beta} E}
                                     }}~.
\label{rewt}
\end{equation}

To further improve sampling, we used  from time to time 
an improved update,
the {\it cut-and-paste move} which is inspired by the Lin-Kernighan
update \cite{LK} for the traveling salesman problem. Here, 
a new polymer conformation
is proposed by randomly choosing a segment of 
$m \le N/2$ consecutive residues, ``removing'' it and ``replacing'' it between
another randomly chosen part of the polymer chain. The energy of the 
new conformation is calculated, and this 
 conformation is accepted or rejected 
according to the
usual Metropolis criterion so that detailed balance is satisfied. The
advantage of the new update is that it changes not only locally a single
dihedral angle, but globally the whole polymer chain and in
this way enhances thermalization. We found that while 
the CPU time increased
moderately by about $\approx 10\%$ the gain in efficiency was much
larger and increased with chain length. It
is obvious that the new update can be used only for simulation of
homopolymers where the number and type of dihedral angles are the same for each
residue.\\

\noindent
{\bf Computational Details}

One MC sweep updates every dihedral angle (in both the backbone and
the side chain) of the poly-alanine chain once. After each MC sweep
we tried once the global {\it ``cut-and-paste''}-update. 

The multicanonical weight factors were determined by the iterative 
procedure described in Refs.~1 and 17. We needed between
40,000 sweeps (for $N=10$)  and 500,000 sweeps (for $N=30$) for their
calculation. All thermodynamic quantities were then calculated from one
production run of $N_{sw}$ Monte Carlo sweeps following additional 
10 000 sweeps
for equilibrization. We chose $N_{sw} = 200,000$ for $N=10$,
$N_{sw}=250,000$ for $N=15$, $N_{sw}=500,000$ for $N=20$, and 
$N_{sw}=1,000,000$ for $N=30$. In all cases,
each simulation started from a completely random initial conformation.

After each fourth sweep we stored the total energy $E_{tot}$, the 
component
energies $E_C$, $E_{LJ}$, $E_{HB}$, $E_{tor}$, the number of helical
residues $n_H$, and the number of helical segments $n_S$ of the 
configuration for further analyses. The latter two quantities
were calculated in the following way: We consider that a residue is
in a (right-handed) $\alpha$-helix configuration when the 
dihedral angles ($\phi,\psi$) fall in
the range ($-70 \pm 20^{\circ},-37 \pm 20^{\circ}$).
The length $\ell$ of a helical segment is then defined by the
number of successive
residues which are in the $\alpha$-helix configuration.
We define a conformation as helical if it has a helical segment with 
$\ell \ge 3$, since a helix  has to have more than
three consecutive residues in a helical state to form the
characteristic hydrogen bond.
The conformation is regarded as completely helical if $\ell \ge N -2$, 
since the two end-residues are flexible. Similarly, a conformation is
considered to be in a coil state if there exists no helical segment 
with $\ell \ge 3$, 
The number $n_H$ of
helical residues in a conformation is defined by the sum of $\ell$
over all $n_S$ helical segments in the conformation.
The same definition was used in our previous work. \cite{OH95a}

The  total number of MC sweeps $N_{sw}$ were chosen so that
the number of ``tunneling events'' $N_{tun} \ge 2$.  A tunneling
event is defined as a series of MC sweeps  in which the poly-alanine
molecule changes from  a completely helical state (lowest-energy
state) to a coil state (high-energy state) and {\bf back}. Hence, 
the value of $N_{tun}$ gives a lower bound on the number of
independent low-energy states found in the simulation.  To further 
quantify
how the numerical effort of the multicanonical simulation of our system
increases with the number of residues, we define 
as ``tunneling time'' $\tau_{tun}$ the average time (in MC sweeps)
needed for a tunneling event and display its values in Tab.~1. 
The  tunneling time $\tau_{tun}$ and therefore the numerical effort
increases with the size of the molecule  as $\tau_{tun} \propto N^x$
where $x=3.1(2)$ when the time is measured in MC sweeps.
  and $x=4.1(2)$
 when measured in updates (since each MC sweep consists of N updates).
While this exponent is large, the numerical performance is still
much better than with a canonical algorithm where one would expect a
supercritical slowing down, i.e. $\tau_{tun} \propto e^{aN}$ ($a$ is an 
 unknown constant).  We remark that the above exponent is close to the optimal
value for the multicanonical algorithm, since the energy grows 
as $E \approx N^2$
with the number of amino acids. Hence, the numerical effort in our simulations
grows like $\approx E^2$ which is what one expects for a random walk
in energy. Similar results were found for spin systems. For these systems
it is also known that 
the numerical effort for calculating the multicanonical weights scales
like $\approx E^{2.5}$. \cite{Berg97} While we did not systematically 
investigate this question, we expect that the same relation holds also
in our case.

\noindent
{\bf RESULTS AND DISCUSSION} \\
In order to elucidate the efficiency of sampling, we first
show in Fig.~1 four conformations obtained during the simulation of
poly-alanine with length $N=30$.  
The four structures
were chosen from a span of half of a tunneling event
that moved  between the 100,000-th MC sweeps and the 156,000-th MC sweep
from a very high-energy state (Fig.~1a) to   a 
lowest-energy conformer (Fig.~1d).  The conformation changes from
a random coil to a completely helical conformation.  One sees that 
our simulation indeed covers a large portion of the conformational space.

In order to study helix-coil transitions in our homopolymers it is necessary
to  first define  an order parameter which allows us to distinguish between
the two phases. Such a quantity can be easily constructed from the
number $n_H$ of helical residues as follows:
\begin{equation}
q = \frac{\tilde{n}_H}{N-2}
\end{equation}
where $\tilde{n}_H$ is the number of helical residues in a conformation,
however, 
without counting the first and last residues. We chose this definition
in order to have $q=1$ for a completely helical conformation. Since 
the residues 
at the end of the polymer chain  can move freely, they will not be
part of a helical segment and therefore should not be counted. On the other 
hand,
$q = 0$ indicates, with our definition, a   coil conformation without any
helical segments. 
We display $<q>_T$ for various
chain length in Fig.~2. The transition between a
high temperature phase characterized by coil states ($<q> \approx 0$)
and a low temperature phase where helical conformations are dominant
($q \approx 1$) is obvious and  becomes sharper with increasing 
length of the poly-alanine chain. 

The helix-coil transition is  correlated
with a change in the average potential energy. This can be seen from 
Fig.~3a-d where we show for all 4 homopolymers  the average total
energy density $<E_{tot}>_T/N$ and the average contributions from the
electrostatic interaction ($<E_C>_T/N$), Lennard-Jones term 
($<E_{LJ}>_T)/N$),
hydrogen-bond energy ($<E_{HB}>_T/N$), and torsion energy ($<E_{tor}>_T/N$).
We display energy densities instead of the energies in order to have 
non-extensive quantities. Again we see in $<E_{tot}>_T/N$ 
 a clear signal for a transition 
 which again becomes more pronounced with increasing  
chain length.  We observe that in the high temperature region 
the quantity $<E_{tot}>/N$ is independent of the length of the chain.
On the other hand, for low temperatures, $<E_{tot}>_T/N$ is a decreasing
function of chain length indicating that here the energy is not
solely a sum of local contributions from each residue. Instead,
the total average energy rather behaves like $<E_{tot}>_T/N  \propto N^a$
with $a \approx 1.4$ (data not shown). 
Examining the results for each component of the potential energy,
 we learn that at small chain length
the change in energy seems to be dominated by the change 
in the Lennard-Jones term. This energy term depends strongly on the
overall size of the molecule and the change in this quantity 
indicates a transition between extended and compact structures.
However, with increasing number of residues, the contribution from the
electrostatic interaction becomes more important indicating the importance
of non-local interactions for the observed transition, although even 
for $N=30$ the Lennard-Jones term is still the most important.

To study the observed transition between helix
and coil conformers in more detail we first  have to determine
 the transition temperature $T_c$. One possibility is to define $T_c$ 
 by the condition that $<q>_{T=T_c} ~= 0.5$. Another possibility 
is to define 
$T_c$ as the temperature
where the change in the order parameter is maximal, i.e., where
$\frac{d}{dT} <q>_T$ has an extremum.  Other sensitive quantities are 
 the susceptibility 
\begin{equation}
\chi_N (T)  =  \frac{1}{N-2} ( <q^2>_T - {<q>_T}^2 )~,
\end{equation}
which is plotted as a function of temperature in Fig.~4, and the 
the specific heat, which we show in Fig.~5. The latter quantity is defined by
\begin{equation}
  C_N (T) \equiv \frac{1}{N~k_B} \frac{d \left( <E_{tot}>_T \right)}{d T}
= {\beta}^2 \ \frac{<E_{tot}^2>_T - {<E_{tot}>_T}^2}{N}~.
\label{eqsh}
\end{equation}
The temperatures where these two quantities have a peak
can again be used to determine the transition temperature.
In Tab.~2 we summarize the estimates for $T_c$ as obtained by the different
methods. For our analyses we divided the $N_{Sw}$ MC sweeps 
in 4  bins and determined $T_c$  in each bin separately. The shown
values are the averages over all bins and the quoted errors were
obtained from the variation of the estimates in the single bins. It is reassuring that the
obtained estimates for the critical temperature do not depend on
the way they were derived. Our final estimate is the average over all
estimates as obtained by the different methods and is 
also displayed in Tab.~2.

We are interested in the transition
temperature for an infinitely long poly-alanine chain. Our data 
for finite chain length show that $T_c$ increases with growing
chain length. By trial and error we found that the 
functional dependence of $T_c$ as
a function of chain length $N$ could be best fitted by 
$T_c (L) = T_c(\infty) - a\cdot e^{-bN}$, which yields  our estimate 
$T_c (\infty) = 514$ K. 

We remark
that we do not find  in Fig.~5 any indications for another
peak in the specific heat at lower temperature $T < T_c$, 
which, if existed,  
could be interpreted as 
a transition between two helix  states. Such a solid-solid 
transition was observed in a recent study on wormlike polymer 
chains.\cite{Jeff}  Similarly, we do not see a shoulder 
in the specific heat for $T >  T_c$. Hence, we conjecture that 
no other transitions but helix-coil one exist for poly-alanine.
Our results
differ here again from the study of the minimal polymer model 
in Ref.~11.
Our data for  the C-peptide
of ribonuclease A  indicates that there may be separate transition
temperatures for
heteropolymers. \cite{HO98c}

The sharp increase in the peak of the susceptibility $\chi_N$ (Fig.~4) and 
the specific heat $C_N$ (Fig.~5) as a function of chain length  suggests
that the observed transition between helix and coil states is a
second-order phase transition.  This assumption is supported by
Fig.~6a where we show the free energy $G(q) = - k_BT\log P(q)$ as a 
function of order parameter $q$ for $T = T_c$ and chain length $N=30$.  
Here, $P(q)$ is the
probability to find a conformation which has the order parameter value
$q$.  We chose a normalization where $G=0$ for  coil state 
($n_H =0$). We see that at the transition temperature the free energy
as a function of order parameter is almost constant for a wide range
of values of the order parameter.  A similar picture holds for the
distribution of potential energy at the transition temperature: states with a 
wide range of energies appear at $T_c$ with similar probability
(data not shown). This is  in support for a second-order phase
transition while for a first-order phase transition
 one would have  a double-peak distribution. 
For comparison we also show in Fig.~6b $G(q)$ for $T=700$ K, which 
is well within  the high-temperature region, and in Fig.~6c $G(q)$ 
for $T=273$ K,
as an example for the behavior of the free energy
in the low-temperature phase. The data are again for our largest chain, 
$N=30$. At high temperatures, coil states are
clearly favored, whereas at low temperature one observes a  strong bias 
towards the total helical state. 

In Fig.~7a and 7b we display the 
average potential energy $<E(q)>$ and entropy $\tilde{S} = k_BT S(q)$, 
respectively, as a function of $q$ 
 at  $T=273$ K ($N=30$). Again we chose a normalization where
$<E(q=0)> = 0$ and $\tilde{S}(q=0) = 0$. We observe for both quantities a
linear decrease with increasing order parameter. These results indicate again 
the existence of long-range order, since we find no indications that
helical segments become unstable once they reach a critical length.
Instead, we  see  in Fig.~8a, where we display the average number of 
helical segments as a function of temperature and chain length, 
that for each chain the low-temperature region is dominated by a single
helix.  Around the critical temperature $T_c$ the number of 
helical segments
is maximal and its average number increases with the size of the chain.
This is again consistent with a second-order phase transition where one
would also expect fluctuations on all length scales at $T_c$. The complementary
picture is shown in  Fig.~8b where we display
the average length of helical segments $\tilde{\ell} = <\ell>/(N-2)$ 
as function of 
temperature.  Our normalization is chosen 
so that the largest possible segments 
 at each chain length will lead to $\tilde{\ell} = 1$. 
We observe that, for low temperatures,  
this quantity is  independent of the number of residues 
which again demonstrates that there is no critical length of helical segments 
in our system  above which helical segments become unstable and break up 
(or this critical length is larger than 30).  For high temperatures
we find that $\tilde{\ell}=<\ell>/(N-2) \to 1/(N-2)$. This is because 
$<\ell> \equiv <n_H>/<n_s>$ is the ratio of the number of helical residues,
divided by the number of helical segments. 
We remark that again neither Fig.~8a nor Fig.~8b gives any indication for
a helix-helix transition as was observed in Ref.~11.  It seems
that the existence of such a transition depends on the chosen model.

According to the Zimm-Bragg model,\cite{ZB}  the average number
of helical residues $<n>$ and the average length $<\ell>$ of a helical
segment are given for large $N$ by
\begin{eqnarray}
{{<n>} \over N}~ &=& ~{1 \over 2} - {{1-s} \over {2 
\sqrt{(1-s)^2 + 4s \sigma}}}~, \\
<\ell>~~ &=& ~1 + {2s \over {1-s+\sqrt{(1-s)^2 +4s \sigma}}}~, 
\end{eqnarray}
where $N$ is the number of residues, and $s$ and $\sigma$ are the
helix propagation parameter and the nucleation parameter, respectively.
Note that $s \ge 1$ implies that ${{<n>} \over N} \ge {1 \over 2}$ 
(more than 50 \% helicity).
 From these equations, with the values of ${<n>} \over N$ and $<\ell>$
calculated from the multicanonical production runs,
one can obtain estimates of $s$ and $\sigma$
parameters. We are here especially interested in the nucleation parameter
$\sigma$ which characterizes the probability for a helix-coil 
 junction, and hence is related to the probability of a helical segment
 breaking apart into two pieces. We show in Fig.~9a the average value
$<\sigma>_T$ as a function of temperature. In accordance with our
previous results in Ref.~1 we notice that $<\sigma>_T$ is
constant below the critical temperature $T_c$ and that its value
in the low temperature region decreases with increasing  chain length.
 From the log-log plot 
(for $T=273$ K, deep in the low-temperature region)
in Fig.~9b  it follows that $<\sigma>$ as a function of chain length
follows a power-law behavior:
\begin{equation}
<\sigma>(N) = \sigma_0 N^{-c}~,     
\end{equation}
(with $\sigma_0 = 0.48(11)$ and $c = 1.13(5)$).
It follows that $<\sigma (\infty)> = 0$:
the probability for a helical segment to break into pieces approaches
zero for infinite chain length. This again supports our claim that 
poly-alanine exhibits a long-range order in the low-temperature phase. 
How can this result be understood 
 in the framework of the Zimm-Bragg
model? The only possibility to have long-range order in a one-dimensional 
Ising model with short interactions is  if the boundary tension 
is infinitely large.
Infinite boundary tension corresponds to $\sigma=0$ in the 
Zimm-Bragg model and describes here a model where only helical conformers
exists (since the probability for helix-coil junctions is
per definition zero in this case). But while in the Zimm-Bragg model $\sigma$ 
is an input parameter and 
describes the interaction between coil and helix segments, 
$\sigma$ is calculated 
in our case {\it a posteriori} and a function of temperature.
Hence, the poly-alanine chains are described by {\it different} 
Zimm-Bragg-models at different temperatures. None of them can exhibit
the phase transition observed for the homopolymer . The existence of this 
phase transition corresponds to the change in  the models.
Finally, in Tab.~3 we summarize the values of $\sigma$ and, in addition,
those of
the helix propagation parameter $s$ at $T=273$ K. The latter quantity
is an increasing function of chain length $N$ and can  be described by 
$s(N) = s_\infty - a e^{-bN}$, which gives $s_\infty = 1.87(3)$.

After having established the existence of a second-order phase 
transition for poly-alanine, we further characterize
this transition by determing the critical exponents.
Conventional arguments for finite-size scaling for a second-order transition
are based on the assumption that the singular part of the free energy
depends only on the system size $N$ and the correlation 
length $\xi$. \cite{FB}
The critical exponents can be extracted from the finite-size
scaling of the heights and width of the peaks in specific heat 
and  susceptibility. With the critical temperature $T_c(N)$  as the 
position where the peak in the specific heat has its maximum, and
$T_1(N)$ and $T_2(N)$ (with $T_1(N) < T_c(N) < T_2(N)$) 
 chosen such that $C(T_1) = 1/2  
C(T_c) = C(T_2)$,  we  have
\begin{equation}
\Gamma_C (N) = T_2(N) - T_1(N) \propto N^{\displaystyle -\frac{1}{\nu}},
\label{nu1}
\end{equation}
and
\begin{equation}
C_N (T_c) \propto  N^{\displaystyle \frac{\alpha}{\nu}}~.
\label{alpha}
\end{equation}
Similarly we find from the heights of the peak in the susceptibility
\begin{equation}
\chi_N(T_c) \propto N^{\displaystyle \frac{\gamma}{\nu}}~,
\label{gamma}
\end{equation}
and from the temperatures where $\chi(T)  = 1/2  \chi (T_c)$
we get a second, independent estimate for the critical exponent $\nu$ by
\begin{equation}
\Gamma_{\chi} (N) = T_2(N) - T_1(N) \propto N^{\displaystyle -
\frac{1}{\nu}}~.
\label{nu2}
\end{equation}
The various quantities are summarized in Tab.~4. From these values
we obtain using the above equations the following estimates for
the critical exponents:
Eq.~(\ref{nu1}) yields  an estimate for $1/\nu = 0.54(5)$ 
(with a goodness  of $q=0.3$ (see Ref.~40 for the 
definition of $q$)
for the fit) which
is comparable with $1/\nu = 0.51(6)$, the value we obtained
with a goodness of $q=0.6$ from the fitting of
 Eq.~(\ref{nu2}). Combining both values, we have
as our final estimate for the correlation length exponent
for the helix-coil transition in poly-alanine:
\begin{equation}
\nu = 1.9 (2)~.
\end{equation}  
With a value of $\alpha/\nu = 0.79(9)$, obtained with goodness
$q=0.6$ by fiting Eq.~(\ref{alpha}), we find the following 
specific heat exponent:
\begin{equation}
\alpha = 1.5 (2)~.
\end{equation}
Similarly, from Eq.~(\ref{gamma}) we obtain, with a goodness
of $q=0.7$, a value $\gamma/\nu = 0.88(7)$, from which we
 get our  estimate for the susceptibility exponent:
\begin{equation}
\gamma= 1.7 (1)~.
\end{equation}
The non-trivial values we obtained for these critical
exponents give further evidence for the second-order
phase transition.

\noindent
{\bf CONCLUSIONS} \\
We have performed multicanonical simulations 
of poly-alanine for
chains of up to 30 residues long
with high statistics.  Our data allowed us to extrapolate
our results to the limit of infinitely long chains and  showed that 
the change
between helical and coil conformers 
can indeed be understood as a temperature-driven, second-order 
phase transition. We were able to determine some critical exponents
for this transition.

\noindent
{\bf Acknowledgment:}\\
Our simulations were  performed on computers
of the Institute for Molecular Science (IMS), Okazaki,
Japan, and the Department of Physics, Michigan Technological University,
Houghton, MI, USA.
This work was supported, in part, by funds from Michigan Technological
University and by a Grant-in-Aid for Scientific Research from the
Japanese Ministry of Education, Science, Sports and Culture. 
\\

\newpage
{\Large Table 1:}\\
\begin{table}[h]
\begin{center}
\begin{tabular}{||l|l||}
$N$ & $\tau_{tun}$ [MC sweeps]\\ 
\hline
10  &  16440 (3498)\\
15  &  22964 (3157)\\
20  & 92872 (31710)\\
30  & 308104 (24254)\\
\end{tabular}
\end{center}
\caption{\baselineskip=0.8cm
           Tunneling times $\tau_{tun}$ measured in MC sweeps as
          obtained from multicanonical simulations of poly-alanine
          with length $N=10, 15, 20,$ and $30$.}
\end{table}

\newpage
{\Large Table 2:}\\
\begin{table}[h]
\begin{center}
\begin{tabular}{||c|c|c|c|c|c||}
$N$  & $<q> = 0.5$ & Max ($\frac{d <q>_T}{d T}$) & Max $ (\chi )$ & 
       Max ($C(T)$)& $T_c$\\
\hline
10 & 432(10) & 422(7) & 430(5)& 423 (7) & 427 (7)\\
15 & 492 (7) & 489 (5)& 495(5)& 490 (5) & 492 (5)\\
20 & 512 (5) & 509 (5)& 513(5)& 510 (5) & 511 (5)\\
30 & 517(10) & 513(10)& 513(10)&513(10) & 513 (10)\\
\end{tabular}
\end{center}
\caption{\baselineskip=0.8cm Estimates for the transition temperature $T_c$ as a
function of the number of residues in the poly-alanine chain.
Shown are the estimates as obtained from the temperature where
 the order parameter $<q>(T) = 0.5$, or where the derivative with
respect to $T$ of $<q>$ has its maximum, or where  the susceptibility
$\chi (T)$ is maximal, or where the specific heat $C(T)$
has an  extremum.  The last column lists our final estimate
for the transition temperature.}
\end{table} 

\newpage
{\Large Table 3:}\\
\begin{table}[h]
\begin{center}
\begin{tabular}{||c|c|c||}
N  &  $\sigma$ & $s$ \\
\hline 
10 &   0.126(4)& 1.561(28)\\
15 &   0.074(2)& 1.679(10)\\
20 &   0.056(1)& 1.780(13)\\
30 &   0.036(1)& 1.845(25)\\
\hline
$\infty$& 0.0    & 1.87(3)  \\
\end{tabular}
\end{center}
\caption{\baselineskip=0.8cm 
Nucleation parameter $\sigma$ and  helix propagation
parameter $s$ as a function of chain length $N$ for $T=273$ K
 as determined from multicanonical simulations. Shown are also our
extrapolated values for an infinitely long chain}.
\end{table}

\newpage
{\Large Table 4:}\\
\begin{table}[h]
\begin{center}
\begin{tabular}{||c|c|c|c|c|c||}
N &  $T_c$ & $C_{max}$ & $\Gamma_{C}$ & $\chi_{max} $ & $\Gamma_{\chi}$\\
\hline
10& 427(7)& 9.3(8) & 146(7) & 0.61(4) & 136(7)\\
15& 492(5)& 12.0(4)& 122(5) &  0.82(3)& 110(5)\\
20& 511(5)& 14.9(6)& 95(5)  &  1.07(4)& 90(5)\\
30& 513(10)& 21.7(1.5)&82(3)&  1.57(8)& 78(3)\\
\end{tabular}
\end{center}
\caption{\baselineskip=0.8cm 
 Maximum of specific heat $C_{max}$ and susceptibility
 $\chi_{max}$ together with width of peak in specific heat
 $\Gamma_{C}$ and width in peak of susceptibility $\Gamma_{\chi}$
 for various chain length.} 
\end{table}

\newpage
{\Large Figures:}\\
\begin{enumerate}
\item Four example configurations chosen from a part of the multicanonical
      simulation where the random walk in energy moved from a high-energy
      state (Fig.~1a) to one of the lowest-energy states (Fig.~1d).
      The figure was created with Molscript \cite{Mol} and
      Raster3D \cite{Ras3d}.
\item Order parameter $<q>_T$ as a function of temperature $T$ for
      poly-alanine molecules of chain length $N=10, 15, 20,$ and $30$.
\item Total energy density $E_{tot}$, electrostatic interaction $E_C$,
      Lennard-Jones term $E_{LJ}$, hydrogen-bond energy density $E_{HB}$,
      and torsion energy density $E_{tor}$ as a function of temperature
      for (a) $N=10$, (b) $N=15$, (c) $N=20$, and (d) $N=30$.
\item Susceptibility $\chi$ as a function of temperature $T$ for 
      poly-alanine molecules of chain length $N=10, 15, 20,$ and $30$.
\item Specific heat $C(T)$ as a function of temperature $T$ for 
      poly-alanine molecules of chain length $N=10, 15, 20,$ and $30$.
\item Free energy $G(q)$ as a function of the order parameter $q$ for
      (a) $T = T_c$, (b) $T=700$ K, and (c) $T = 273$ K. We chose the
      normalization $G(0) = 0$. All results are for  $(Ala)_{30}$.
\item Average potential energy $<E>(q)$ (a) and entropy $\tilde{S}(q)
      = k_BT S(q)$ (b) as a function of order parameter $q$  at
      temperature $T = 273$ K. We chose the normalization $<E>(q=0) = 0$
      and $\tilde{S}(q=0) = 0$. All results are for $(Ala)_{30}$.
\item Average number of helical segments $n_S$ (a) and average length
      of helical segments $\ell/(N-2)$ (b) as a function of temperature $T$ for
      poly-alanine molecules of chain length $N=10, 15, 20,$ and $30$.
\item (a) Nucleation parameter $\sigma$ as a function of temperature $T$ for
       poly-alanine molecules of chain length $N=10, 15, 20,$ and $30$.
      (b) Log-log-plot of the nucleation parameter $\sigma$ as a
      function of the number of residues at $T=273$ K.
\end{enumerate}
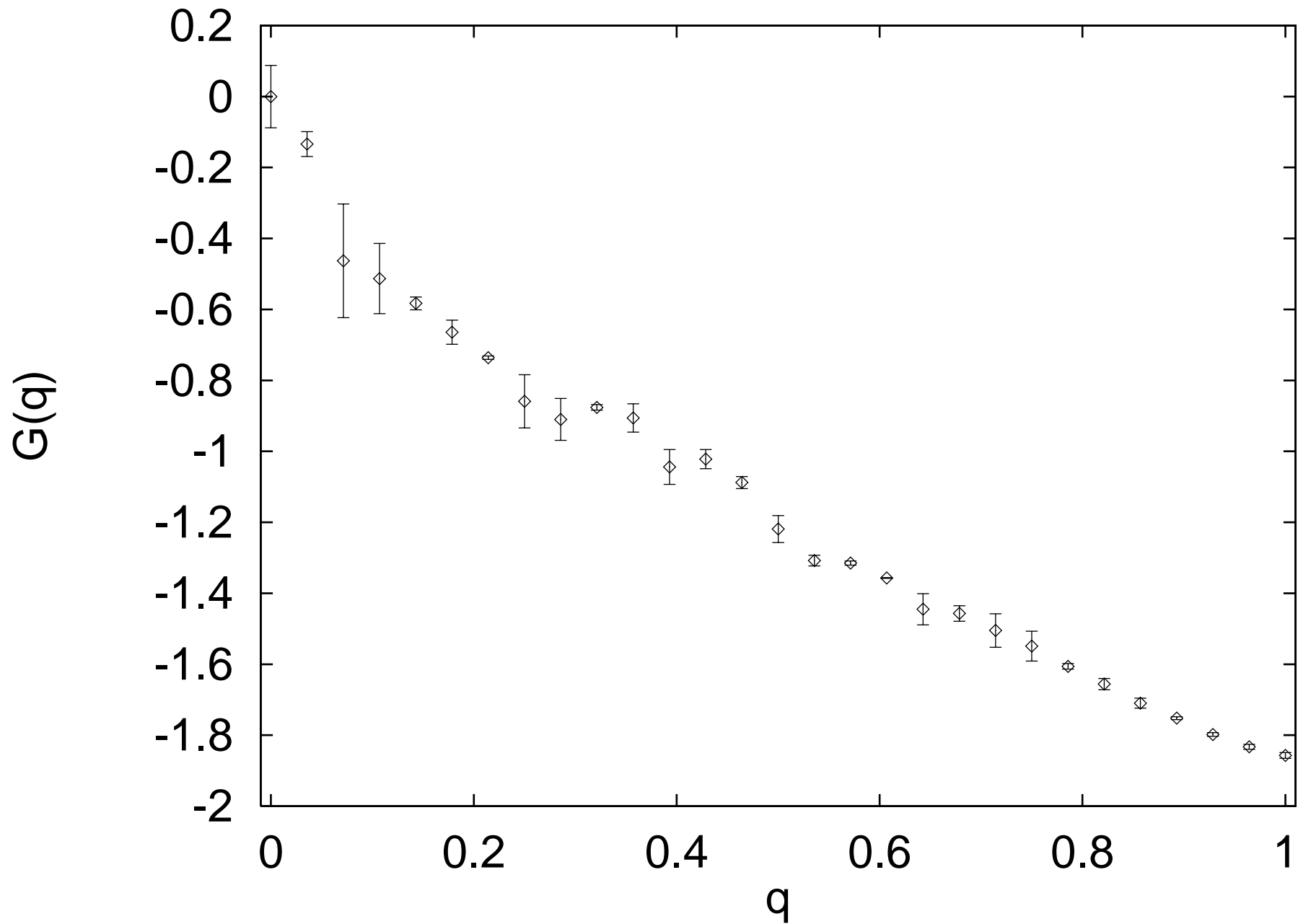
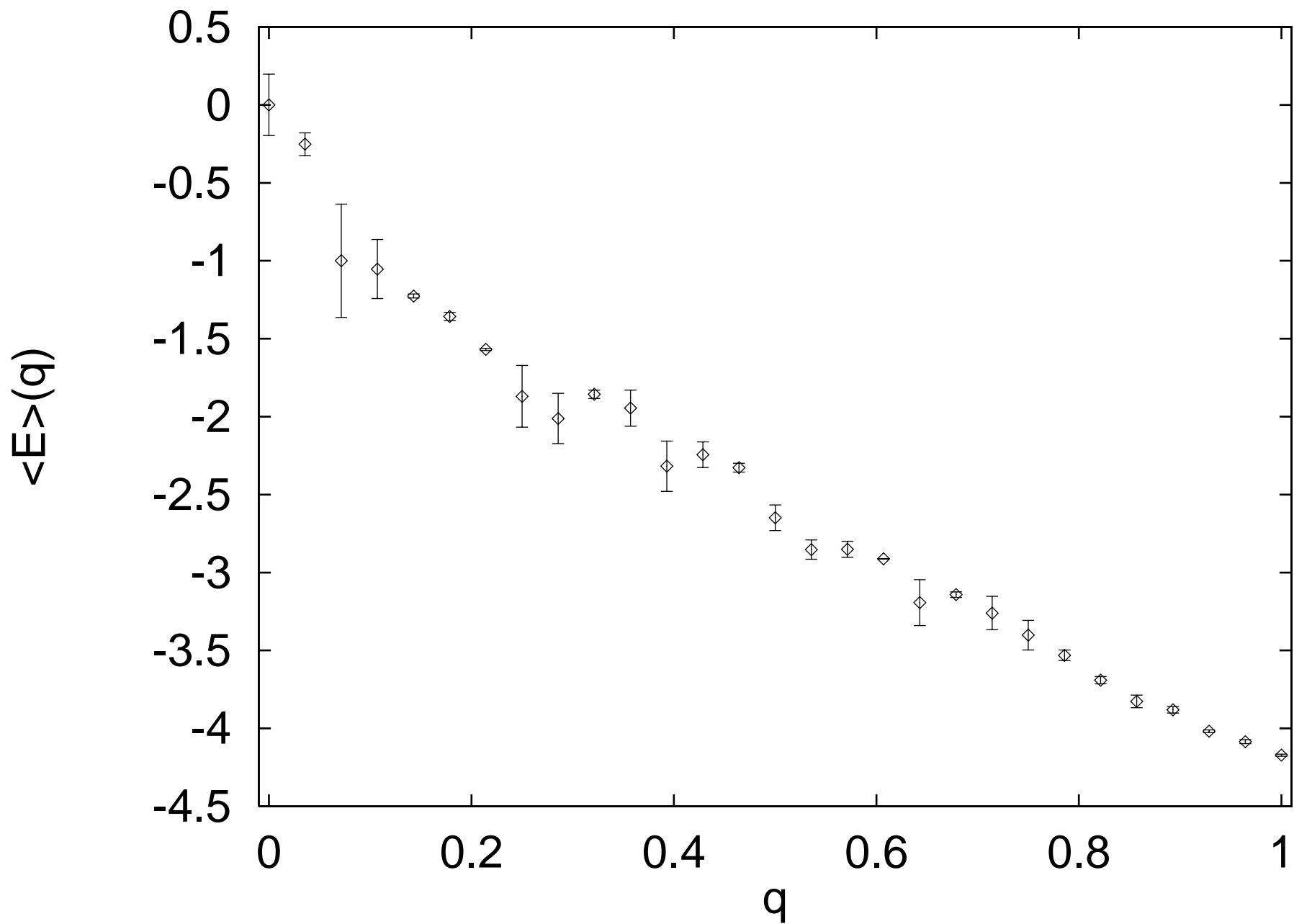
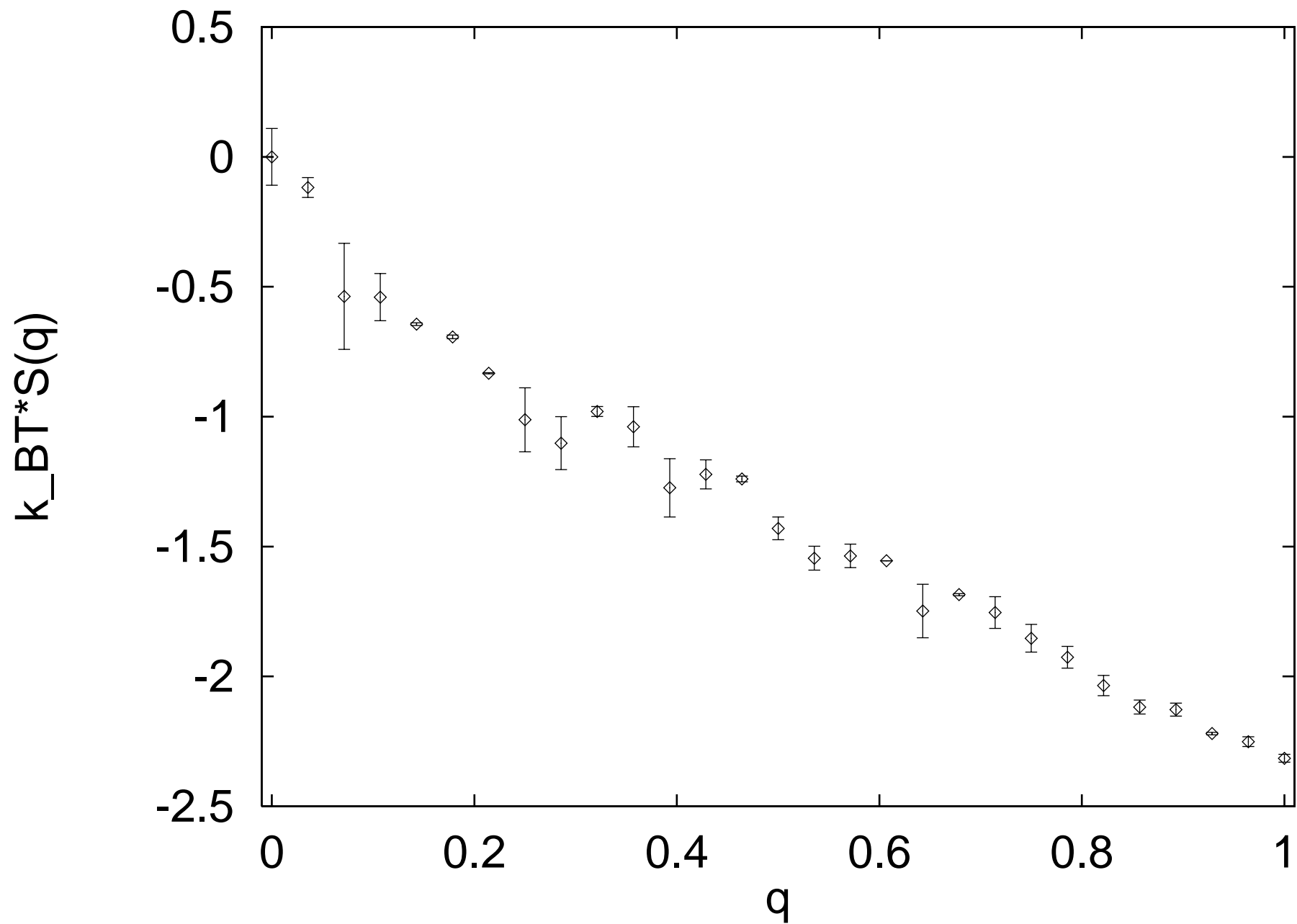
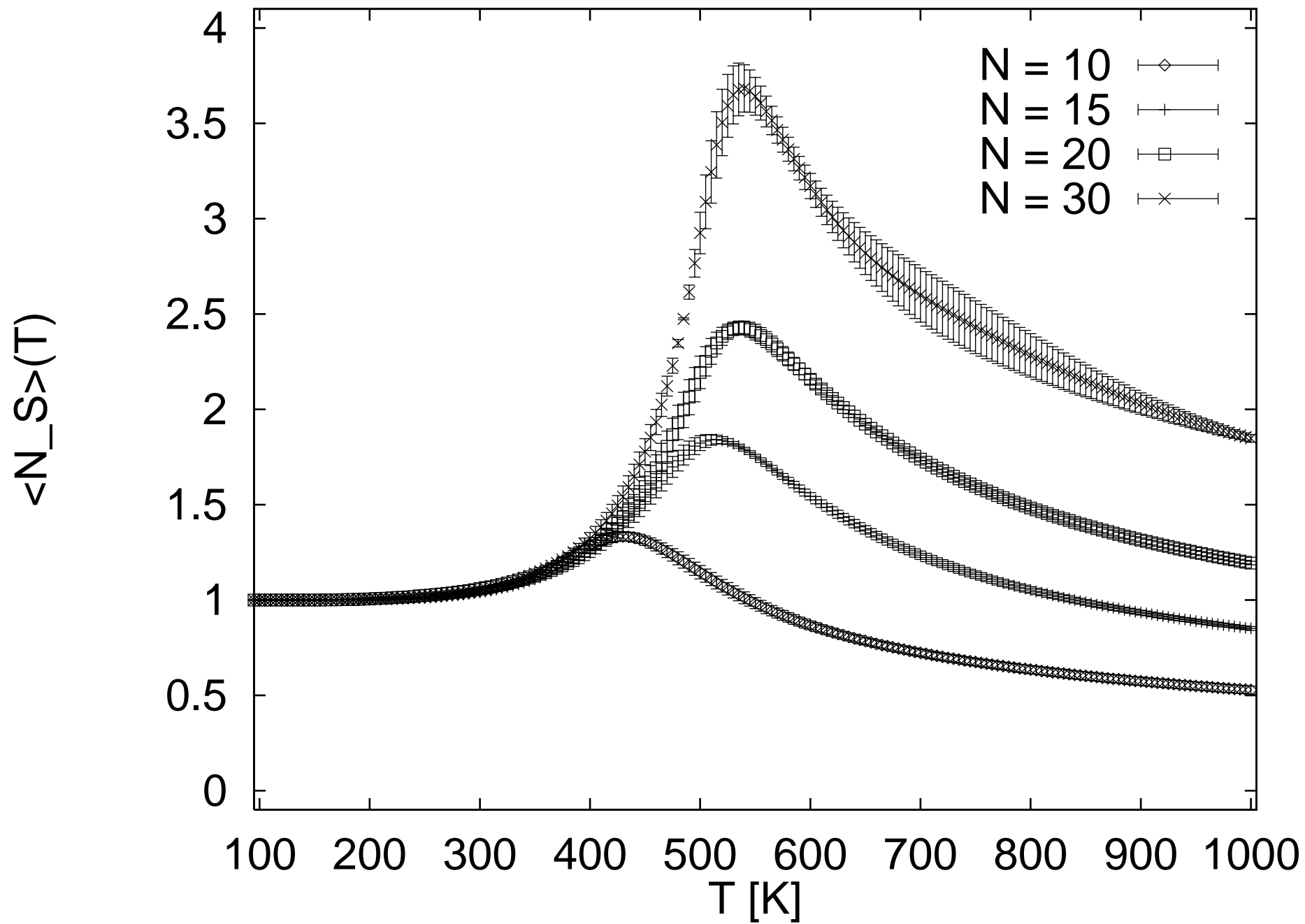
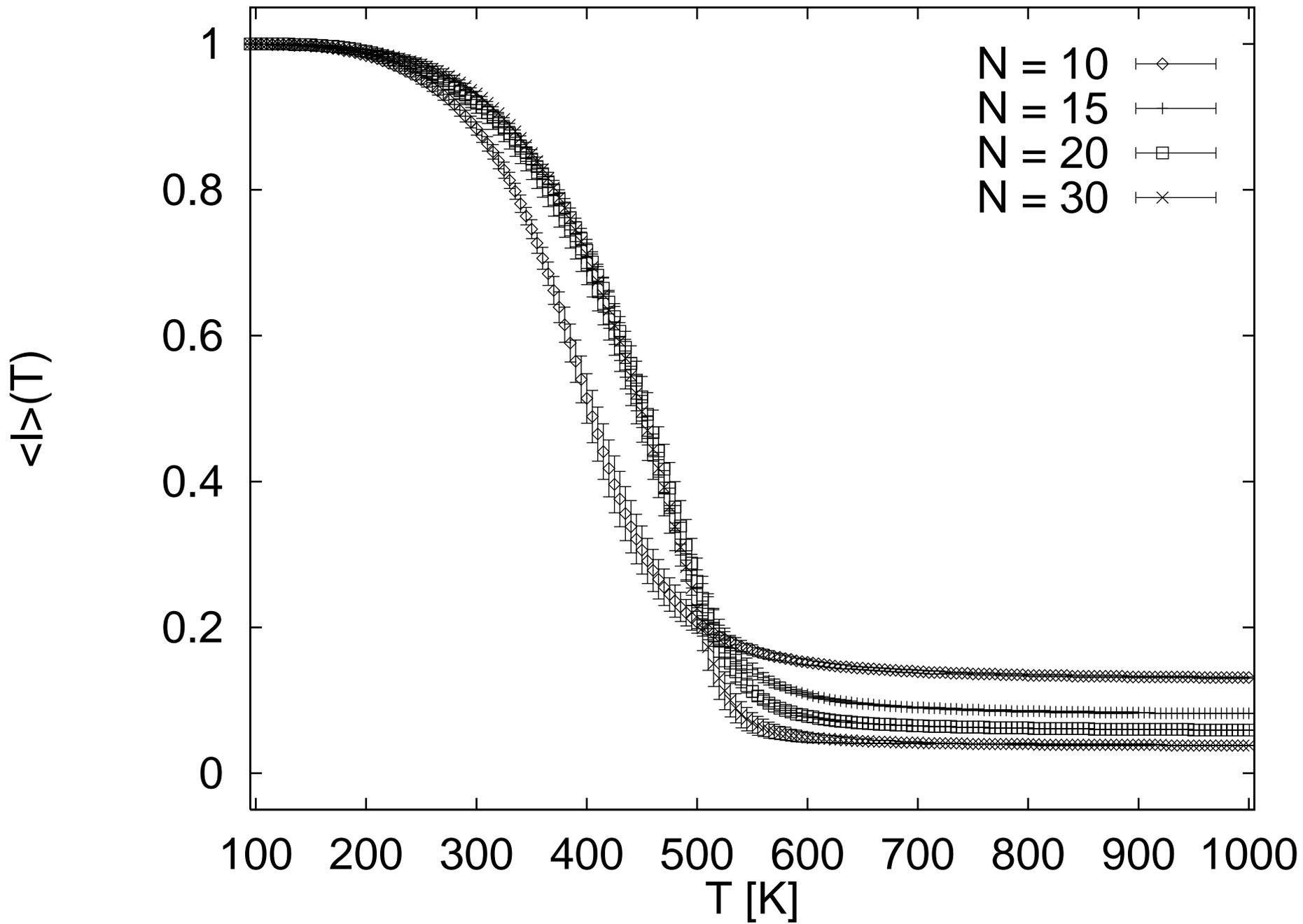
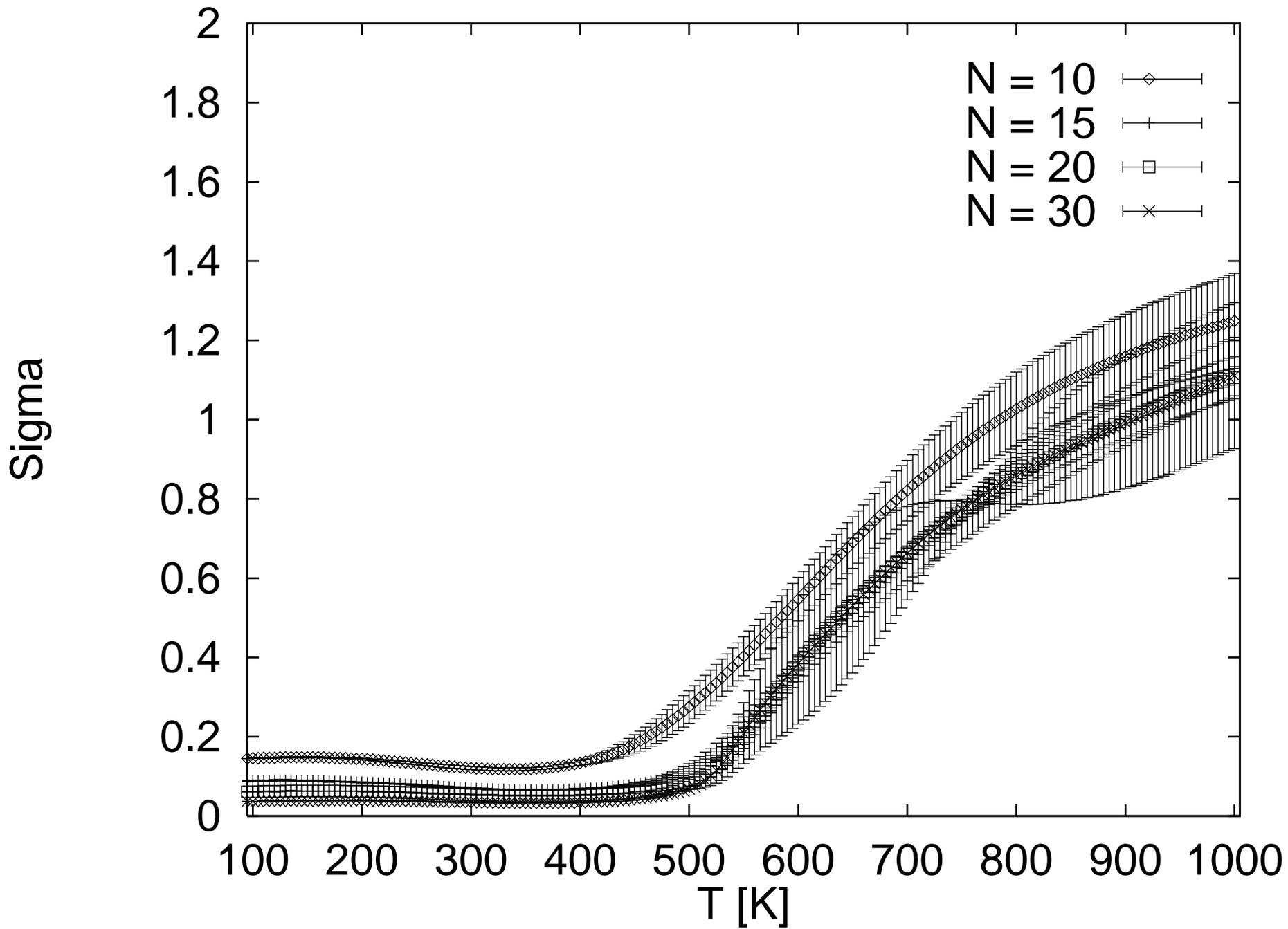
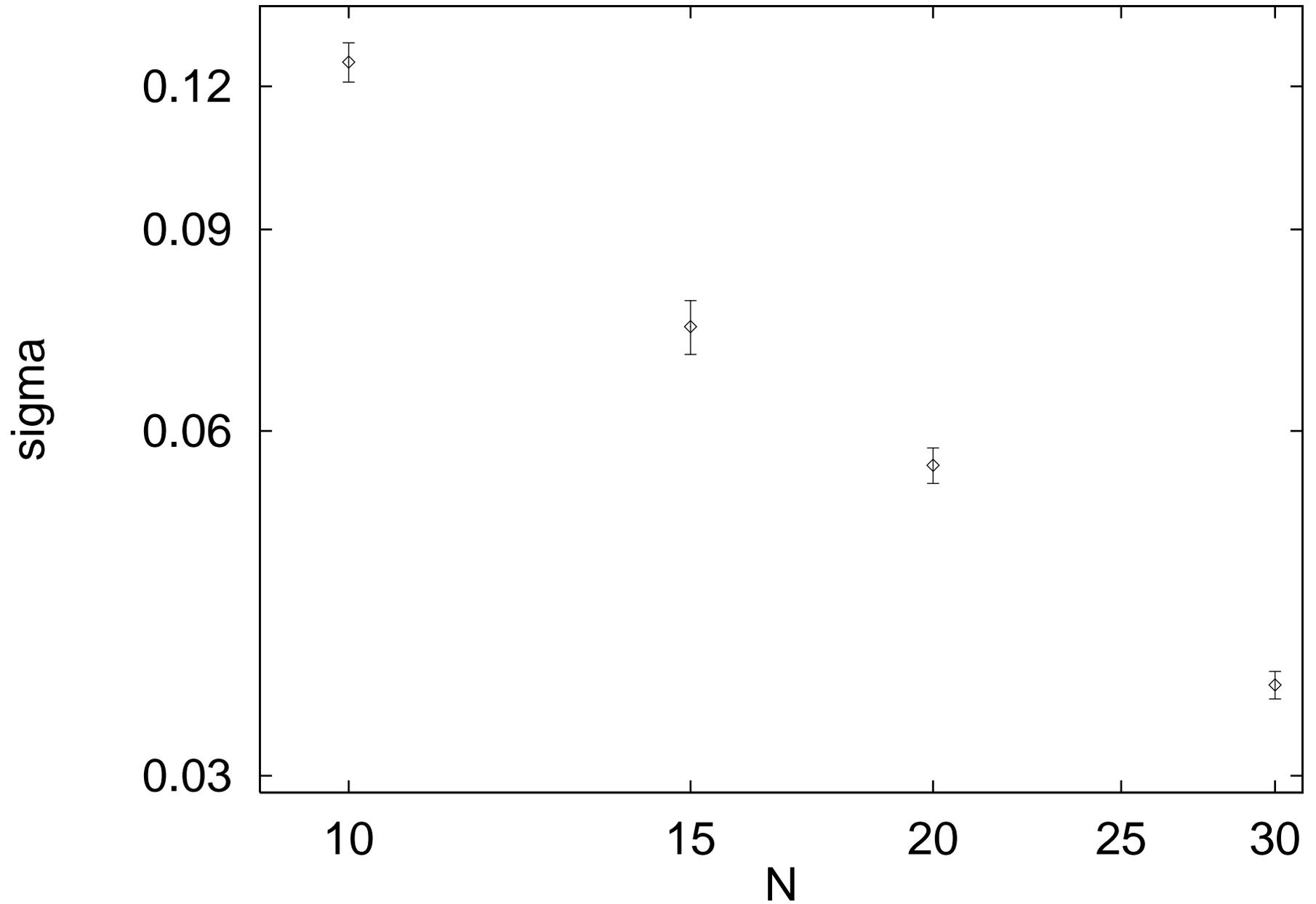
\end{document}